\documentclass[usenatbib]{mn2e}

\usepackage{amsmath}
\usepackage{graphicx}
\usepackage{epstopdf}
\usepackage{multicol}

\title{Microlensing as a possible probe of event-horizon structure in quasars}
\author[Tomozeiu et al]{Mihai Tomozeiu\thanks{mihai@physik.uzh.ch},$^{1,2}$ 
Irshad Mohammed,$^{1,2,3}$ 
Manuel Rabold,$^{1,2}$
Prasenjit Saha,$^{1,2}$  
\newauthor
and Joachim Wambsganss$^{1,4}$\\
$^1${Physik-Institut, University of Zurich, Winterthurerstrasse 190,
  8057 Zurich, Switzerland} \\
$^2${Institute for Computational Science, University of Zurich,
  Winterthurerstrasse 190, 8057 Zurich, Switzerland} \\
$^3${Theoretical Astrophysics Group, Fermi National Accelerator Laboratory, Batavia, IL 60510, USA}\\
$^4${Zentrum f\"ur Astronomie der Universit\"at Heidelberg,
  M\"onchhofstrasse 12--14, 69120 Heidelberg, Germany}
}

\begin{document}

\maketitle

\begin{abstract}

In quasars which are lensed by galaxies, the point-like images sometimes show sharp and uncorrelated brightness 
variations (microlensing). These brightness changes are associated with the innermost region of the quasar passing 
through a complicated pattern of caustics produced by the stars in the lensing galaxy. In this paper, we study 
whether the universal properties of optical caustics could enable extraction of shape information about the central 
engine of quasars. We present a toy model with a crescent-shaped source crossing a fold caustic. The silhouette 
of a black hole over an accretion disk tends to produce roughly crescent sources. When a crescent-shaped source 
crosses a fold caustic, the resulting light curve is noticeably different from the case of a circular luminosity 
profile or Gaussian source. With good enough monitoring data, the crescent parameters, apart from one degeneracy, 
can be recovered.

\end{abstract}

\begin{keywords}
Supermassive black holes, microlensing, quasars.
\end{keywords}

\section{introduction}

Active galactic nuclei are thought to be powered by the
accretion of matter from the proximal environment into a supermassive
black hole.  The radiation emitted excites the surrounding medium
which becomes detectable as narrow line regions, broad line regions
and optical continuum.  Moreover, in the direction perpendicular to
the accretion disc, where the medium is more transparent, jets will
appear.  If a jet is oriented towards the Earth, a quasar is observed
\citep[e.g.,][]{1984RvMP...56..255B}.  While the basic mechanism
\citep[originating in the work
  of][]{1964ApJ...140..796S,1964SPhD....9..246Z,1969Natur.223..690L}
is not in doubt, the central engines, near the event horizons of the
black holes, remain to be probed.

For the black holes in the Galactic centre and the centre of M87
---the two nearest objects that barely qualify as AGN--- the central
engines (which are $<0.1\rm\,mas$ on the sky) are close to being
resolved through very long baseline interferometry, which shows
preliminary indications of the jet-launching structures
\citep{2008JPhCS.131a2055D,2012Sci...338..355D,2013MNRAS.434..765K,2016arXiv160205527F}.
Current data do not deliver images but require fitting to
predefined models for the images.  A whole range of models have been
applied, starting from simple geometric models to more complex
physical models
\citep{2008Natur.455...78D,2011ApJ...738...38B,2009ApJ...706..497M,2010ApJ...717.1092D}.
The more complicated models nonetheless tend to predict a crescent
shaped silhouette of the black hole.  This motivated
\cite{2013MNRAS.434..765K} to use a simple geometric crescent model to
fit the data, and argued that the crescent is nothing but the
silhouette of the event horizon.

The great majority of quasars, however, lie at redshifts beyond 2
\citep{2014A&A...563A..54P} and their central engines would be orders
of magnitude smaller on the sky. The direct observations of the black
hole silhouettes of quasars are far beyond foreseeable instrumentation.
In the present paper, we consider a possible indirect method, related
to \cite{1999ApJ...524...49A}, through which gravitational
microlensing could probe the black hole shadow and its proximal quasar
environment.

Gravitational microlensing of quasars reviewed in
Section~\ref{sec:microlensing} below, refers to sharp changes in the
observed brightness of quasars that have been lensed by an intervening
galaxies, without any changes in the intrinsic luminosity.
Microlensing affects only the light from the innermost part of the
quasar, such as the optical continuum
\citep[e.g.,][]{2012A&A...544A..62S} and is a consequence of two
things: the very small size of the central engine, and granularity of
the mass distribution of a lensing galaxy due to stars.  The latter
means that the local magnification is not a smooth function of source
position. It contains a complicated network of singular curves, known
as caustics.  Figure~\ref{fig:magnification_map} shows part of a
magnification map with a few caustics.  The lensed brightness would be
given by placing the source on such a magnification map and
integrating the surface brightness weighted by the magnification.
Most astrophysical sources straddle several caustics, and hence, their
net brightness varies smoothly with location.  The central engine of
quasars, however, is smaller than the typical spacing between
caustics.  As a result, the lensed brightness undergoes sudden changes
as a quasar crosses a caustic.  The effect supplies an upper limit on
the size of the central engine, and can also be used to study the mass
distribution and kinematics of stars in the lensing galaxy as well
\citep[e.g.,][]{2012ApJ...744..111P}.  Caustics have an additional
remarkable feature: though they can be very complicated, they have
some universal properties well-known from catastrophe theory.  In
particular, very close to the simplest caustics (known as folds), the
magnification is approximately constant on one side and
$\propto1/\sqrt p$ where $p$ is the transverse distance of the source
from the caustic.  This property will be exploited later.

In Section~\ref{sec:source-models} we introduce the three source
profiles used in our subsequent models and simulations: a
constant-brightness disc, a circular Gaussian, and the crescent source
introduced by \citep{2013MNRAS.434..765K}.  The latter is simply a
constant-brightness disc with a smaller, non-concentric disc cut out
of it.  We also derive the half-light radius for a crescent.  The
half-light radius can characterise the source size for all three types
of source.

Section~\ref{sec:fold-crossing} shows the light curves that result
when each of the model sources crosses an ideal fold.  This would
apply in Figure~\ref{fig:magnification_map} to sources along the path
AB or BC, for sources small enough that the curvature of the caustics
is negligible.  With this assumption one can imagine the caustic as an
infinite wall to be crossed by the source as presented in
Figure~\ref{fig:infinite_fold}.  The source brightness distribution
parallel to the caustic naturally makes no difference to the
observable brightness; each source can be replaced by an effective
one-dimensional source profile, by flattening the source so it becomes
perpendicular to the caustic.  In principle, the effective
one-dimensional brightness profile could be recovered from the light
curve by deconvolution.  \cite{1999ApJ...524...49A} modelled this
profile as the result of a circular accretion disc seen through the
spacetime around a Kerr black hole, and \cite{2012MNRAS.423..676A}
have applied the idea to observed light curves to infer properties of
quasar accretion discs.  In this work, we take a simpler but arguably
more robust approach: we study features in the light curves
characteristic of a crescent-like source which in turn would indicate
a black-hole silhouette.  Figure~\ref{fig:char_points} shows the
qualitative features: there is a period during which the dark cutout
disc is crossing the caustic, and before and after there are periods
when the only the bright parts of the crescent are in transit across
the caustic.  The details depend on the orientation of the crescent,
but basically the dark disc causes a rising light curve to plateau or
dip.  These features are still present, albeit faintly, if the simple
crescent is replaced by a source based on an accretion-disc simulation
of a black-hole environment
(Figures~\ref{fig:M87_image}--\ref{fig:M87_plots}).

In Section~\ref{sec:numerics} we carry out source fitting to
lightcurves, with both noise and systematic errors are present.  We
generate three lightcurves by taking a uniform disc, a circular
Gaussian, and a crescent across the path AB in
Figure~\ref{fig:magnification_map}, and then adding noise.  The path
simulates crossing a clean but not ideal fold.  In addition, we
generate templates lightcurves by running the three source types, with
various parameter values, across the path CB.  That is, the templates
come from a similar but not identical caustic, thus deliberately
generating a systematic error.  We then fit the noisy lightcurves to
the templates using Markov chain Monte-Carlo.  We find that the
correct source type can be inferred from the $\chi^2$ values.  The
parameter values can also be inferred. The fitting errors are larger
than the formal uncertainties, which is expected in the presence of
systematic errors, but still appear acceptable.

Finally in Section~\ref{sec:discussion} we discuss in more depth the
implications of the results presented in the previously mentioned
sections.  One of the most interesting implication is the possibility
to estimate the black hole mass from the reconstructed
parameters. This further requires good approximations of the relative
transversal velocity between the quasar and the lens. Proper
constraints can be set with independent observations of the stellar
structure that contain the gravitational lens.

The thorough study of the possibility to reconstruct the quasar's
structural parameters from light-curves containing multiple
microlensing events represents the target of future work. This will
most probably require the use of powerful statistical tools. Another
path for future work is the designing of an observation regime best
suited for acquiring the necessary microlensed light-curve.
\section{Microlensing}\label{sec:microlensing}

We start by introducing in a succinct manner the gravitational lensing
theory that is relevant for microlensing in general and for the scope
of the present paper in particular.  More detailed presentation of the
theory can be found in several references, such as
\cite{2001stgl.book.....P}.

\subsection{Magnification}

The gravitational lens equation
\begin{equation}
\vec\beta = \vec\theta - \vec\alpha(\vec\theta)
\label{eqn:lens}
\end{equation}
relates the apparent sky position $\vec\theta$ of a light source to
its true but unobservable sky position $\vec\beta$ through the bending
angle $\vec\alpha$.  The latter is an integral 
\begin{equation}
\vec\alpha = (1+z_{_L})\frac{D_S-D_L}{D_SD_L} \frac{4G}{c^2}
\int \Sigma(\vec\theta')
\frac{\vec\theta-\vec\theta'}{|\vec\theta-\vec\theta'|^2}\,
d^2\vec\theta'
\label{eqn:alpha}
\end{equation}
depending on the projected density $\Sigma(\vec\theta)$ (kg/steradian)
of the lens, the lens redshift $z_{_L}$ and the comoving distances
$D_L$ and $D_S$ to the lens and source.  The derivative of the
apparent position with respect to the source position
\begin{equation}
M(\vec\theta) =
\left(\frac{\partial\beta}{\partial\theta}\right)^{-1}
\label{eqn:magnif-matrix}
\end{equation}
is known as the magnification matrix, and its determinant
\begin{equation}
\mu(\vec\theta) = \det|M(\vec\theta)|
\end{equation}
is the brightness amplification of an image of a point source.  In
other words, the source will brighten or dim according to whether
$\mu(\vec\theta)$ is more or less than unity.  If there are multiple
mages at distinct but not observationally resolved $\vec\theta_i$ from
the same $\vec\beta$, a total brightness amplification
of\begin{equation} \mu_{\rm total} = \sum_{i} \mu(\vec\theta_i)
\end{equation}
applies. It is possible for the magnification to become formally
infinite, as a result of an eigenvalues of the magnification matrix
(\ref{eqn:magnif-matrix}) becoming infinite.  This generically happens
on curves on the $\vec\theta$ plane, known as {\em critical curves}.
Mapping a critical curve to the source plane, through the lens
equation (\ref{eqn:lens}), to the source plane $\vec\beta$ gives the
so-called caustics.  Caustics can appear in the optical system, not just
gravitational lensing.  For a point source, caustics are singularities
of the magnification; for finite-size sources caustics correspond to
high and sharply changing magnification.

Caustics are important in all forms of lensing with multiple images,
but they have a special significance for lens quasars, first pointed
out by \cite{1979Natur.282..561C}.  The granularity of the mass
distribution $\Sigma(\vec\theta)$ due to individual stars produces a
caustic network on the scale of $\sqrt{GM_\odot D_L/c^2}$ or $10^{-6}
\sqrt{M/M_\odot}\rm\,arcsec$ for typical lens and source redshift
\citep{2001PASA...18..207W}.  Extended sources wash out this
micro-caustic structure, but the optical-continuum source of quasars
is even smaller.

\begin{figure}
\centering
\includegraphics[width=0.9\hsize]{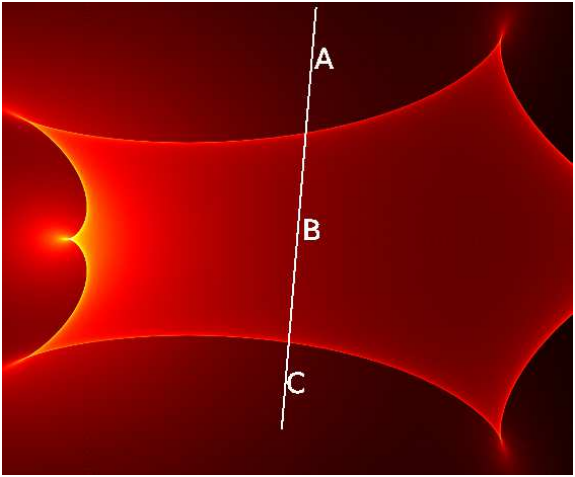}
\caption{\label{fig:magnification_map} Magnification map used later in
  this paper (Section~\ref{sec:numerics}). The white line marks the
  trajectory of the center of the sources for which the lightcurves
  presented in figure 11 are generated.  Note that this represents an
  atypically simple region of any realistic magnification map.}
\end{figure}

\subsection{Magnification near a fold caustic}

A simple example of a caustic network is shown in
Figure~\ref{fig:magnification_map}.  There are two general categories
of caustics in gravitational lensing, cusps and fold, and examples of
both kinds can be seen in this figure.  Magnification near a caustic
has universal properties, independent of the system and has been
extensively studied
\citep{1986ApJ...310..568B,1992A&A...260....1S,2002ApJ...574..970G,2002ApJ...580..468G}.
In particular, at distance $p$ from a fold caustic
\begin {equation}
 \mu(p) = \mu_0 + C_0 \frac{1}{\sqrt{p}} \Theta(p).
\end {equation}
Here the magnification of a point source near a caustic is equal to
the sum of the magnification due to other reasons $\mu_0$, assumed to
be locally constant, and a decrease with the square root of the
distance from the fold. The latter term becomes activated only after
the source enters the region interior to the caustic curves when the
values of the step function $\Theta(p)$ become unity. The
proportionality constant $C_0$ depends on the local conditions in the
vicinity of the caustic.

A source of arbitrary shape can be described by a two-dimensional brightness function $S_{2D}(p - p_s, q - q_s)$ defined for a coordinate system $p,q$ where $p_s, q_s$ denote the coordinates of the center of a source.

\begin{figure}
\includegraphics[width = .49\textwidth]{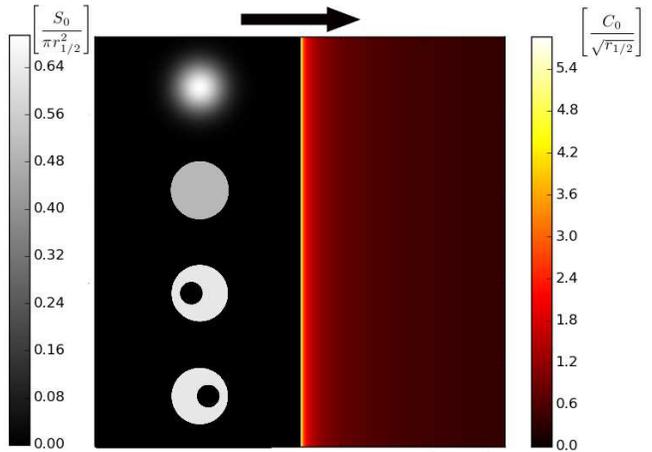}
\caption{\label{fig:infinite_fold} Source profiles for Gaussian, disk,
  crescents (right) and magnification map for an infinite fold
  (left). Objects have the same $S_{0}$ and $r_{1/2}$. The black arrow marks the direction of motion of the
sources relative to the caustic.}
\end{figure}

For a microlensing event the lightcurve can be written for an
undefined source shape as:
\begin{equation}
 F(t) = \int_{-\infty}^\infty \int_{-\infty}^\infty S_{2D}(p-p_s(t), q-q_s(t)) \mu_t(p) \mathrm{d}q \mathrm{d}p
 \label{eqn:ft2d}
\end{equation}
In order to build the previous equation we have considered that the
time dependency of the flux $F$ is given only by the motion of the
source with respect to a fixed caustic. Therefore the only time
dependent quantities in the right hand side of the equation are the
coordinates of the center of the source $p_s,q_s$ and by construct
$S_{2D}$.

Due to the choice of the coordinate system the amplification factor has no dependence on the $q$ coordinate. The previous equation can be rewritten as:

\begin{equation}
 F(t) 
= \int_{-\infty}^\infty  \mu_t(p) S_{1D}\left(p-p_s(t)\right) \mathrm{d}p,
\label{eqn:ft}
\end{equation}
\\
where we have defined the one dimensional flux function as:
\begin{equation}
 S_{1D}(p-p_s(t)) = \int_{-\infty}^\infty S_{2D}(p-p_s(t), q-q_s(t)) \mathrm{d}q
\end{equation}

This representation is a valid approximation only when the apparent size of the source is much smaller than the corresponding Einstein angle of the lens. In this context 
all the information about the source shape and brightness that can be contained in the lightcurve is exhaustively given by the 1D flux function.
In other words, if two sources with different $S_{2D}$ have the same $S_{1D}$ they cannot be distinguished by studying their lightcurves.

\section{Models for extended sources}\label{sec:source-models}

In the present study we are analysing three types of sources with different surface brightness: 
\begin{enumerate}
 \renewcommand{\theenumi}{(\arabic{enumi})}
  \item a rotationally symmetric source with a bivariate gaussian surface brightness distribution,
  \item a disk source with constant surface brightness distribution,
  \item a crescent shaped source with constant surface brightness distribution.
\end{enumerate}
The first two sources are the typical choices used in the literature to describe the luminous parts of a quasar (Prasenjit should give some citations here). The third one is a recently proposed
variant \citep{2013MNRAS.434..765K}.

\subsection{Rotationally symmetric source with a bivariate gaussian surface brightness distribution}

A symmetric 2D gaussian can be described mathematically as:

\begin{equation}
 S_{2D}^G(p-p_s, q-q_s) = \frac{S_0^G}{2 \pi \sigma^2} e^{-\frac{(p-p_s)^2}{2 \sigma^2}} e^{-\frac{(q-q_s)^2}{2 \sigma^2}}.
\end{equation}
\\
The corresponding 1D brightness is:

\begin{equation}
 S_{1D}^G(p-p_s) = \frac{S_0^G}{\sqrt{2 \pi} \sigma} e^{-\frac{(p-p_s)^2}{2 \sigma^2}}.
\end{equation}
\\
Other parameters of the model are the total flux $S_0^G$ and $\sigma$. 

Although such a definition for a source would have non-zero surface/linear brightness for any coordinate $p,q$, the amount of light received by a detector from outside a $3 \sigma$ disk centered at $p_s, q_s$ 
would be insignificant. For a gaussian distributed surface brightness source the half-light radius is directly proportional to the parameter $\sigma$ according to the equation:
\begin{equation}
r_{1/2} = \sqrt{ln(4)} \sigma.
\end{equation}

\subsection{Disk source with constant surface brightness distribution}

One can construct mathematically a disk source with constant surface brightness and radius $R$ using a stepfunction:
\begin{equation}
 S_{2D}^D(p-p_s, q-q_s) = \frac{S_0^D}{\pi R^2} \Theta \left( R^2 - \left( p-p_s \right)^2 - \left( q-q_s \right)^2 \right).
\end{equation}
By integrating over $q$ coordinate the linear brightness function is obtained:

\begin{equation}
 S_{1D}^D(p-p_s) = \frac{2 S_0^D}{\pi R}  \sqrt{1 - \frac{(p-p_s)^2}{R^2} }    \Theta \left( R^2 - \left( p-p_s \right)^2 \right).
\end{equation}
\\
The half-light radius of a uniform disk source is $R/\sqrt{2}$.

\subsection{Crescent source with constant surface brightness distribution}\label{subsec:crescent}

The surface brightness distribution of a geometric crescent can be built by considering two disk sources of constant brightness. One larger disk will contribute positively to the total flux while one smaller disk 
that is interior to the large one will contribute negatively. This superposition can be written for 2D as:\\

\begin{equation}
 S_{2D}^C =  S_{2D}^{Dp} -  S_{2D}^{Dn}  
 \label{eqn:s2d}
\end{equation}
with\\

\begin{equation}
 S_{2D}^{Dp}(p-p_{sn}, q-q_{sn}) = \frac{S_0^{Dp}}{\pi R_p^2} \Theta \left( R_p^2 - \left( p-p_{sp} \right)^2 - \left( q-q_{sp} \right)^2 \right)
\end{equation}
\\
and
\begin{equation}
 S_{2D}^{Dn}(p-p_{sn}, q-q_{sn}) = \frac{S_0^{Dn}}{\pi R_n^2} \Theta \left( R_n^2 - \left( p-p_{sn} \right)^2 - \left( q-q_{sn} \right)^2 \right).
\end{equation}
\\
The following notations were used: $R_p, (p_{sp}, q_{sp}), R_n, (p_{sn},q_{sn})$ are the radii and coordinate of the center for the larger positive disk and smaller negative disk respectively.  $S_0^{Dp},S_0^{Dn}$ represent the total flux of radiation received from the large and small disk. From this point forward we will not use the total flux from each source. Instead we will 
use the difference which in this case is th total flux from the crescent-shaped source $S_0^C$. \\
Equation \ref{eqn:s2d} can be written as:\\

\begin{align}
 S_{2D}^C &= \frac{S_0^C}{\pi \left(R_p^2-R_n^2 \right)} \left\{ \Theta \left[ R_p^2 - \left( p-p_{sp} \right)^2 - \left( q-q_{sp} \right)^2 \right] \right.\nonumber\\
 &\qquad \left. {} -  \Theta \left[ R_n^2 - \left( p-p_{sn} \right)^2 - \left( q-q_{sn} \right)^2 \right] \right\}.
\end{align}
\\
Analogous for the linear brightness function:

\begin{align}
 S_{1D}^C &= \frac{2 S_0^C}{\pi \left(R_p^2-R_n^2 \right)} \left\{ \sqrt{R_p^2 - (p-p_{sp})^2}  \Theta \left[ R_p^2 - \left( p-p_{sp} \right)^2 \right] \right.\nonumber\\
 &\qquad \left. {} - \sqrt{R_n^2 - (p-p_{sn})^2 } \Theta \left[ R_n^2 - \left( p-p_{sn} \right)^2 \right] \right\}.
\label{eqn:s1_d}
\end{align}

There are some constraints on the parameters used to define a crescent in the previously presented manner that need to be stated. First, we must impose the obvious $R_p > R_n$ relation. Secondly, 
the small disk must always be interior to the large disk:
\begin{equation}
 R_p \ge R_n + \sqrt{\left(p_{sp} - p_{sn} \right)^2 + \left(q_{sp} - q_{sn} \right)^2}
\end{equation}

For the distances between the centers of the two disks we will use the same notations as the one found in the paper \citep{2013MNRAS.434..765K}, $a \equiv p_{sn} - p_{sp}$ and $b \equiv q_{sn} - q_{sp}$.

The half-light radius of any source is invariant to any rotational transformation. In the present case of a crescent source the effective radius is dependent on the parameters $R_p$, $R_n$ and $\sqrt{a^2+b^2}\equiv c $ exclusively. From symmetry considerations the centroid of the source is collinear with the centers of the two disks and it is situated at a distance $d_c$ from the center of the bright disk. $d_c$ can be computed numerically with the use of a variation of equation \ref{eqn:s1_d}:

\begin{equation}
\begin{aligned}
\frac{S_0^C}{2} & = \frac{2 S_0^C}{\pi \left(R_p^2-R_n^2 \right)} \int_{d_c}^{R_p} \bigg[ \sqrt{R_p^2 - p^2} \\ 
        & - \sqrt{R_p^2 - \left(p-c\right)^2} \Theta \left(R_p^2 - \left(p-c\right)^2 \right) \bigg] dp. 
\end{aligned}
\end{equation}   

With the position of the centroid determined, the half-light radius can be also be computed numerically:
 
\begin{equation}
\begin{aligned}
S_0^C &=  \frac{4S_0^C}{\pi \left(R_p^2-R_n^2 \right)} \int_{0}^{r_{1/2}} \left[ \Theta_2  - \Theta_1  \right] p dp; \\
\Theta_1 &= \bigg[ \pi - \arccos \frac{R_p^2 + p^2 -d_c^2}{2 R_p p} \\
         & \phantom{= \bigg[ \pi} - \arccos \frac{R_p^2 - p^2 + d_c^2}{2 R_p d_c}
            \bigg] \Theta \left( p - R_p + d_c \right); \\
\Theta_2 &=  \pi - \arccos \frac{ p^2  + (c+d_c)^2 - R_n^2}
                                {2 p \left(c + d_c \right)} \\
         & \phantom{= \pi - } \Theta \left( p + R_n - d_c -c \right). 
\end{aligned}
\end{equation}

\begin{figure}
\includegraphics[width = .49\textwidth]{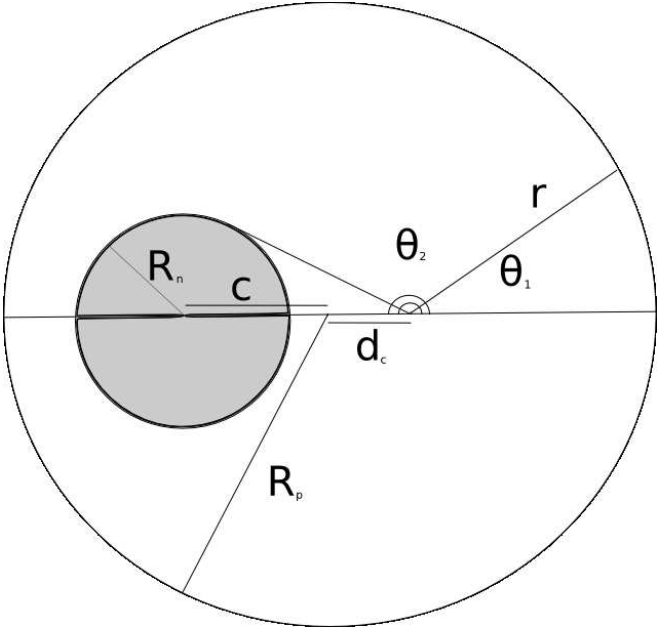}
\caption{\label{fig:geom_crescent} Geometry of crescent sources.}
\end{figure}

\section{Lightcurves of the extended sources during fold crossing}\label{sec:fold-crossing}

Using equation \ref{eqn:ft} and the one-dimensional flux function
presented in the previous section one can compute numerically the
lightcurves of the three extended sources for the simplified
infinite-wall-caustics model.

\subsection{Lightcurve of the gaussian source}

The amount of light received by an observer from a source with a gaussian distributed brightness with $\sigma$ and total flux $S_0^G$ in the absence of any gravitational lensing is:
\begin{equation}
 F^G(t) = \int_{-\infty}^\infty  \left( \mu_0 + \frac{C_0}{\sqrt{p}} \Theta \left( p \right) \right) \left( \frac{S_0^G}{\sqrt{2 \pi} \sigma} e^{-\frac{(p-p_s(t))^2}{2 \sigma^2}} \right) \mathrm{d}p.
\end{equation}

which can be simplified to:
\begin{equation}
 F^G(t) = \mu_0 S_0^G + \frac{C_0 S_0^G}{\sqrt{2\pi} \sigma} \int_{0}^\infty \frac{e^{-\frac{(p-p_s(t))^2}{2 \sigma^2}}}{\sqrt{p}} \mathrm{d}p.
\end{equation}

\begin{figure}
\centering
    \includegraphics[width = 0.48\textwidth]{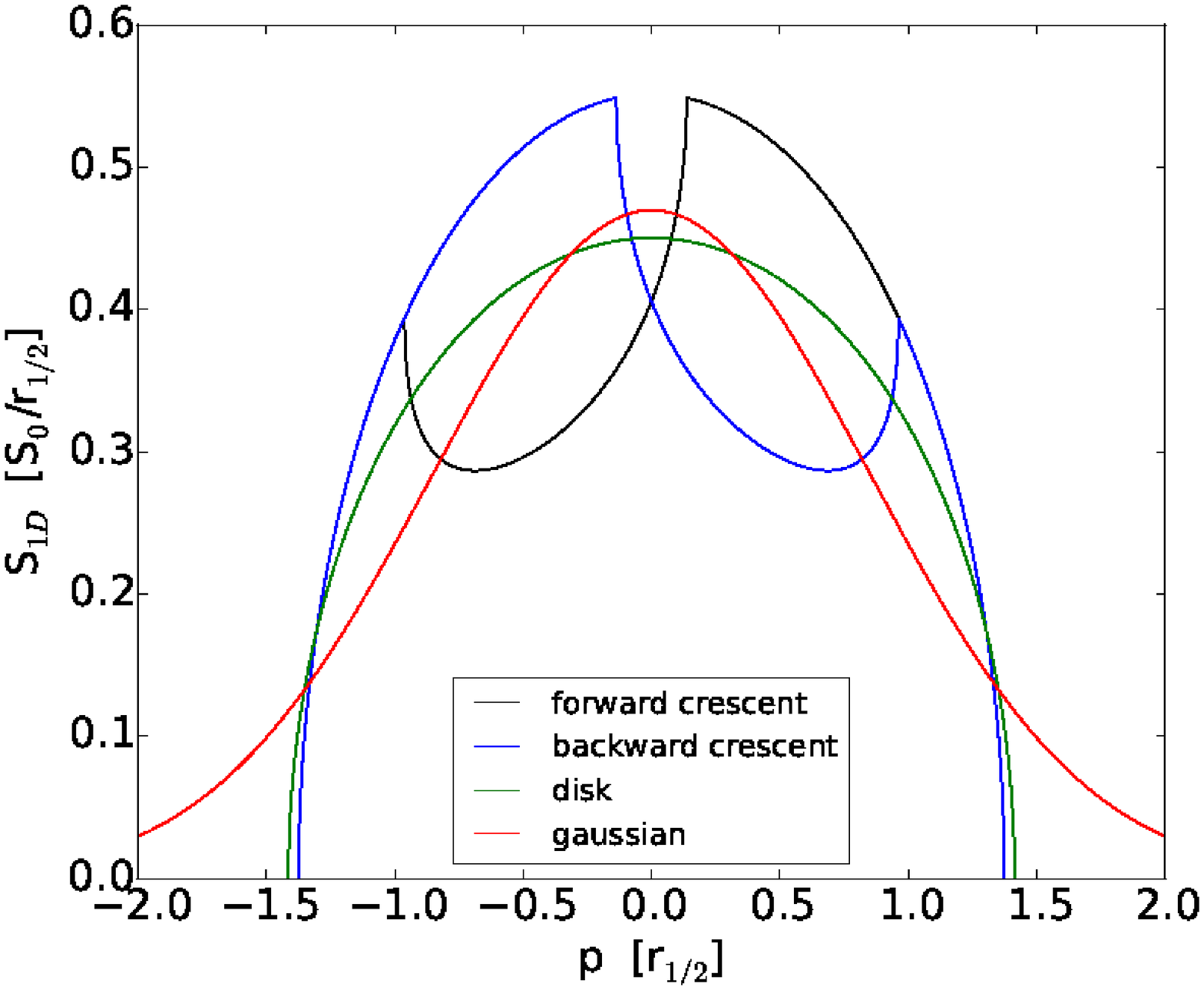}
    \includegraphics[width = 0.48\textwidth]{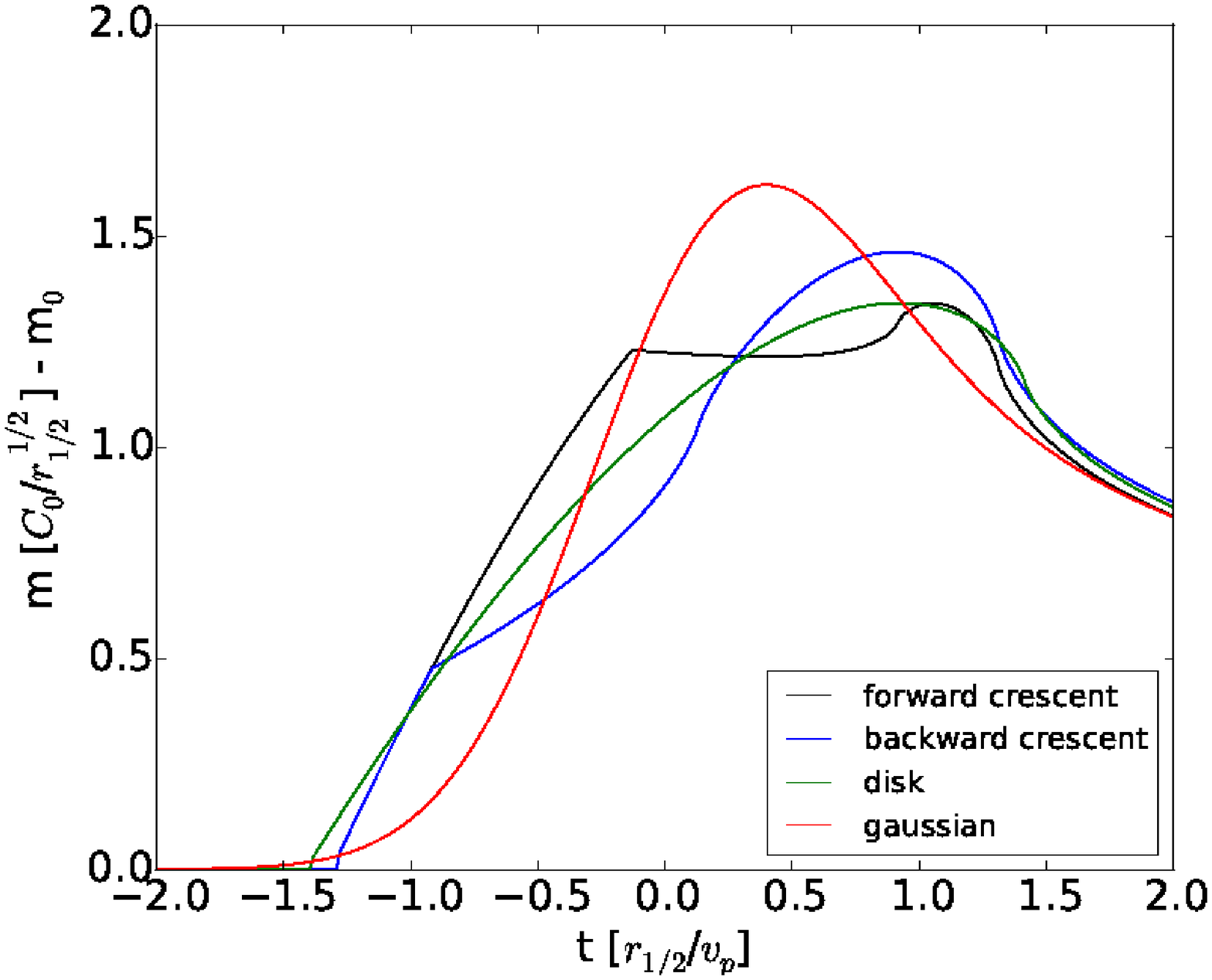}
\caption{\label{fig:lightcurve_gauss} Lightcurves of crescents, Gaussian and disk for sources with identical $S_0$ and $r_{1/2}$. Crescent source has $r_{1/2}$ = 0.72754 $R_p$, $R_n$ = 0.4 $R_p$, a = 0.3/-0.3 $R_p$ }
\end{figure}

\subsection{Lightcurve of the disk shaped source}

Analogous to the gaussian shaped source, the disk source with uniform brightness, radius $R$ and unmagnified flux $S_0^D$ has a lightcurve described by the equation:

\begin{equation}
\begin{aligned}
 F^D(t) &= \int_{-\infty}^\infty  \left( \mu_0 + \frac{C_0}{\sqrt{p}} \Theta \left( p \right) \right) \\
    & \bigg[ \frac{2 S_0^D}{ \pi R} \sqrt{1 - \frac{\left( p-p_s(t) \right)^2}{R^2}}  \Theta \left(R^2 - \left(p-p_s(t) \right)^2 \right) \bigg] \mathrm{d}p.
\end{aligned}
\end{equation}

which is equivalent to:
\begin{equation}
 F^D(t) = \mu_0 S_0^D + \frac{2 C_0 S_0^D}{\pi R} \int_{max(0, p_s(t) - R)}^{max(0, p_s(t) + R)} \frac{1}{\sqrt{p}} \sqrt{1 - \frac{\left( p-p_s(t) \right)^2}{R^2}} \mathrm{d}p.
\end{equation}

\begin{figure}
\centering
    \includegraphics[width = 0.48\textwidth]{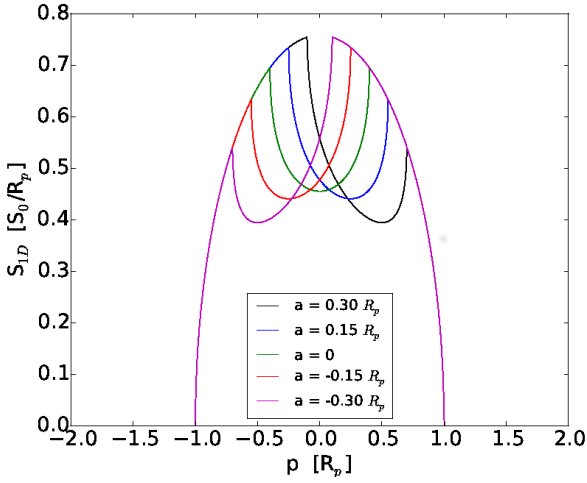}
    \includegraphics[width = 0.48\textwidth]{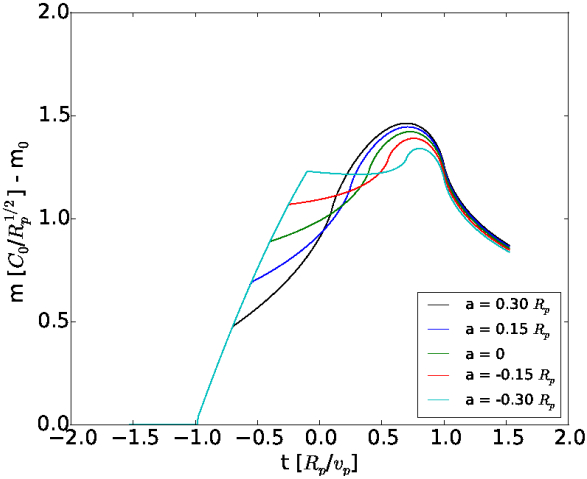}
\caption{\label{fig:lightcurve_disk} Lightcurves of crescent sources with $R_n$ = 0.4 $R_p$.}
\end{figure}

\subsection{Lightcurve of the crescent shaped source}

The lightcurve of a crescent shaped source with unamplified flux $S_0^C$, radii $R_p$, $R_n$ and center displacement $a(t)$ is:

\begin{equation}
\begin{aligned}
 F^c(t) &= \mu_0 S_0^C + C_0 \frac{2 S_0^C}{\pi \left( R_p^2 -R_n^2 \right) } \\
    &\bigg[ \int_{max(0, p_s(t) - R)}^{max(0, p_s(t) + R)} \sqrt{\frac{R_p^2 - \left( p-p_s(t) \right)^2 }{p}} \mathrm{d}p \\
    &  -  \int_{max(0, p_s(t) - a(t) - R)}^{max(0, p_s(t) -a(t) + R)} \sqrt{\frac{R_p^2 - \left( p-p_s(t) +a(t) \right)^2 }{p}} \mathrm{d}p  \bigg] .
\end{aligned}
\end{equation}

\begin{figure}
\centering
    \includegraphics[width = 0.48\textwidth]{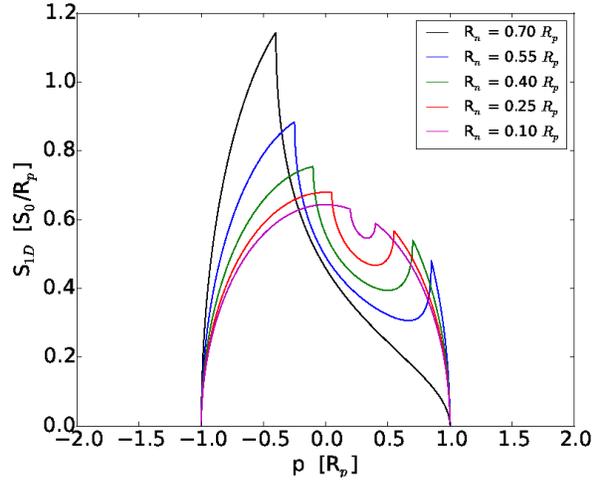}
    \includegraphics[width = 0.48\textwidth]{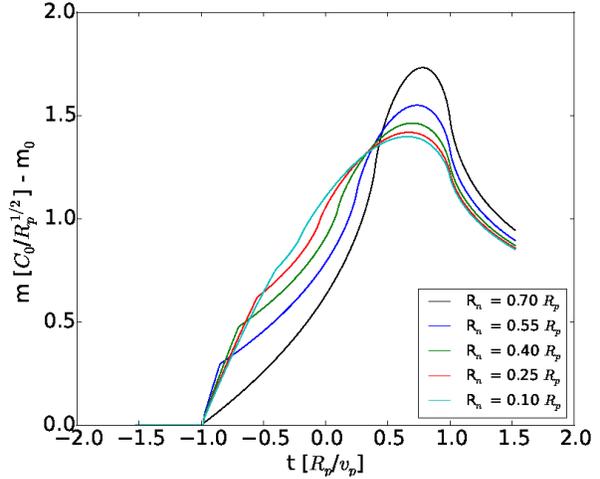}
\caption{\label{fig:lightcurve_crescent_back} Lightcurves of backwards crescents with  a = 0.3 $R_p$.}
\end{figure}

The function $p_s(t)$ can be chosen to be equal to $v_p(t-t_0) + p_{s0}$. Where $p_{s0}$ is the coordinate $p$ of the source at the initial time, and $v_p$ is the component of the velocity
along the $p$ axis. Such a modelling of the motion of the object in the source plane describes a linear motion with constant velocity. Furthermore, we reduce the complexity of the model by 
choosing the function $a(t)$ to be constant in time.  

\begin{figure}
\centering
    \includegraphics[width = 0.48\textwidth]{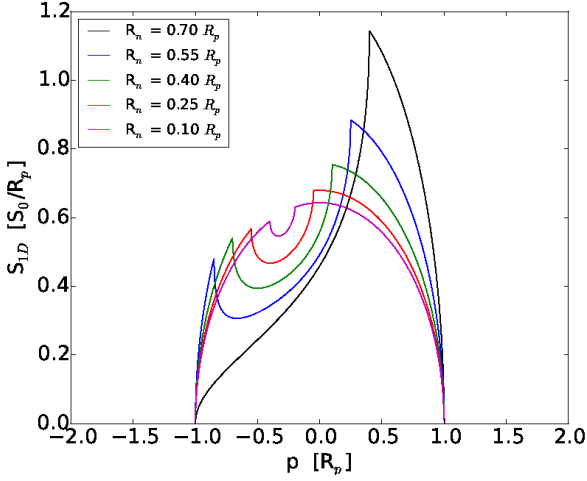}
    \includegraphics[width = 0.48\textwidth]{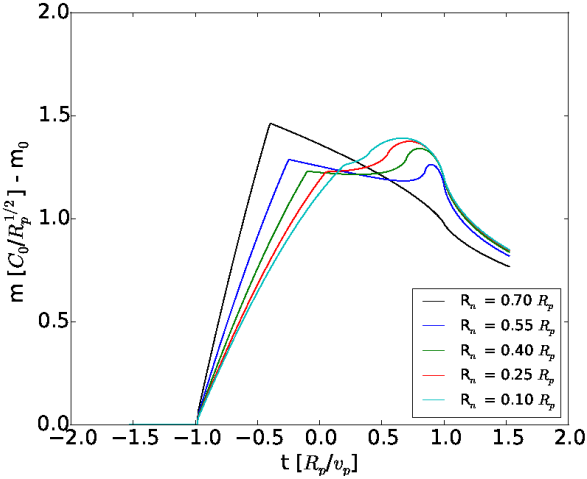}
\caption{\label{fig:lightcurve_crescent} Lightcurve of forwards crescents with a = -0.3 $R_p$.}
\end{figure}

There are four characteristic points visible on the resulting one-dimensional light profile of the crescent shaped source. Two of the points, p1 and p4, mark the outer boundaries of the luminous disc component. The other two, p2 and p3, mark the boundaries of the dark disc component. Since $S_{1D}$ is a projection of $S_{2D}$ on a line perpendicular to the caustic, the following relation holds:
\begin{equation}
    p_2-p_1 = R_p -R_n - a.
\end{equation}
In addition there are two other obvious relations the characteristic points which are independent of the projection:
\begin{equation}
    p_4 -p_1 = 2 R_p,
\end{equation}

\begin{equation}
        p_3 -p_2 = 2 R_n.
\end{equation}
All four points mark the positions where derivative $\frac{dS_{1D}}{dp}$ is discontinuous. The points can be used to define three regions: $p_1 - p_2$ where $S_1D$ is convex, $p_2 - p_3$ where $S_1D$ is concave, and $p_3 - p_4$ where $S_1D$ is convex again. \\

Due to the nature of the caustic and the monotonic behaviour of the magnification map on both sides of the caustic. The previously mentioned characteristic points are inherited by the microlensing lightcurve. The points on the temporal dimension $t_1, t_2, t_3$ and $t_4$ correspond to instances in time when the fold is aligned with $p_1, p_2, p_3$ and $p_4$, respectively. For a constant relative velocity $v_p$ between the source and the caustic there is a simple relation between the points $p_i$ and instances $t_i$:

\begin{equation}
    t_j - t_i = \frac{p_j - p_i}{v_p}.
\end{equation}

With the use of the previous four equations the following identities can be written:

\begin{equation}
    R_p = \frac{v_p \left( t_4 -t_1 \right)}{2}, 
\end{equation}

\begin{equation}
        R_n = \frac{v_p \left( t_3 -t_2 \right)}{2}, 
\end{equation}

\begin{equation}
        a = \frac{v_p \left( t_4 +t_1 - t_3 - t_2 \right)}{2}. 
\end{equation}

\begin{figure}
\centering
	\includegraphics[width = 0.48\textwidth]{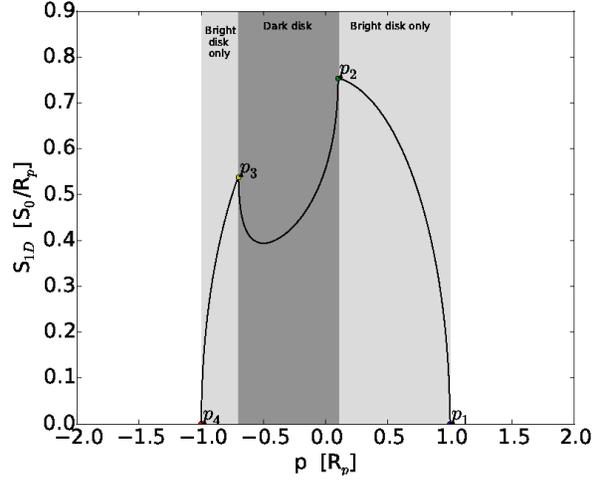}
        \includegraphics[width = 0.48\textwidth]{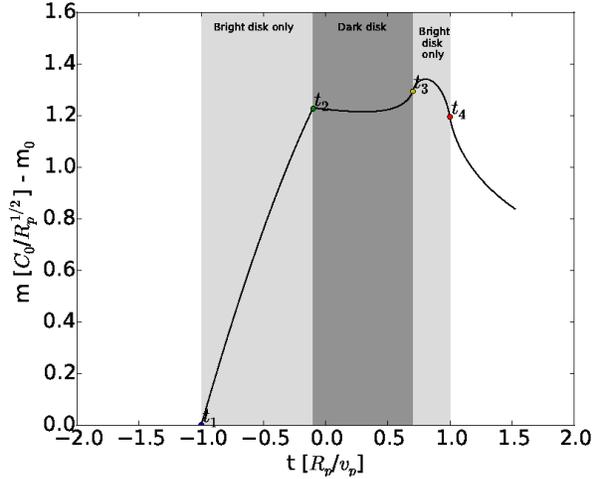}
\caption{\label{fig:char_points} The characteristic points and instances of a crescent source's one dimensional profile and microlensing lightcurve. }
\end{figure}

Figure 4 reveals that the lightcurve of a crescent source has more visible features than the other two light-curves
corresponding to the disc and gaussian shape. There are three regions where $S_{1D}$ and by inheritance, the lightcurve
 has distinguishable behaviour. The first region that would be recorded on a lightcurve plot represents the period
of time when the bright disk begins to be overlapped by the caustic and stops when just before the dark disk reaches
the caustic. During this period of time, the flux of light from the source is increasingly magnified. The second
period starts and ends with the overlapping of the dark disc. As a boundary of the two regions, there is a
distinguishable point where the slope of the magnification is drastically changed. This apparent discontinuity in the
first derivative of the magnification function is caused by the caustic amplification of the sudden drop in the $S_{1d}$
function. During the respective period, the magnification growth slows or even reverses during the first part of the
period and starts to grow faster again as the dark disk ends its overlap with the caustic. At the point where
the dark disc clears the caustic, the growth of the magnification is infinite, which appears as a saddle point
on the lightcurve. Next, the final period corresponds to the case when the dark disc has cleared the caustic
and the bright disc continues to overlap with the caustic. During this period, the lightcurve reaches a peak
that for most of the parameter space is global and for the rest of the parameter space local.
At the end of the period, the growth of the magnification is negative infinite. the respective point appears 
as a second saddle point on the lightcurve. Past this point the magnification of any finite source will 
decay in roughly the same manner. \\

The impact of parameter $a$ on the shape of the lightcurve can be observed in figure 5 (a-var). Sources where 
the center of the dark disc reaches the caustic before the center of the bright disc are characterized by a smoother
broad peak in contrast to the cases where the center of the dark disc reaches the caustic after the bright disk.
In the case of the latter the instance when the dark disc reaches the caustic corresponds to larger and
larger magnifications until it becomes a local and even global peak. The effect of the $R_n$  parameter on the
light curve is presented in figures 6 and 7. For the particular set of parameters where $R_p = R_n +a$ the third
period of time discussed previously does not exist. Particularizing further, if the value of the radius of the dark
disc is comparable to the value of the radius of the bright one then the position and shape of the maximum magnification
 are strikingly different. In case the crescent reaches the caustic with the bright region first, the peak magnification
 happens when the dark region reaches the caustic and it is characterized by a sharp variation in magnification growth,
In the opposite case, the peak appears before the end of the bright disc reaches the caustic and shape is smoother. \\

\subsection{Microlensing a simulated image of M87}

\cite{2012MNRAS.421.1517D} have created a radiative image of M87 based on the GRMHD simulations presented in \citep{2009MNRAS.394L.126M}. 
The top right-most image in figure 5 of \cite{2012MNRAS.421.1517D} has been projected to a 1D profile associated to the perpendicular 
direction to a fold caustic approaching the image from the right. The projection is presented in the upper panel of figure 9. 
The amplification values of the flux of light corresponding to a microlensing event are presented in the lower panel of figure 9. 
In general, the behaviour of the lightcurve is similar to the geometric crescent source with the caveat that the outer regions 
surrounding the luminous parts of the image have non-zero flux and thus are more extended than the simplified source model.

\begin{figure}
\centering
    \includegraphics[width = 0.48\textwidth]{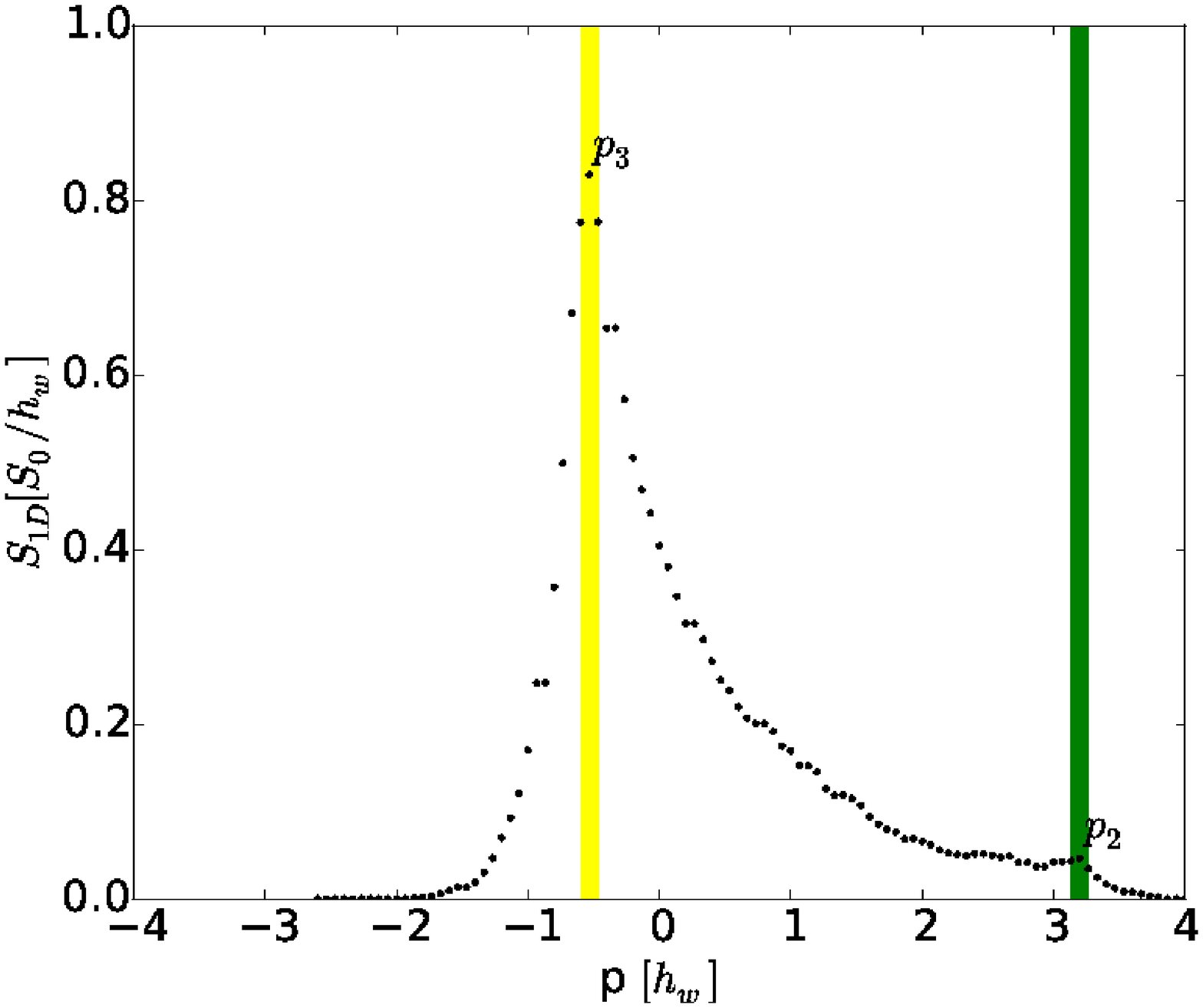}
        \includegraphics[width = 0.48\textwidth]{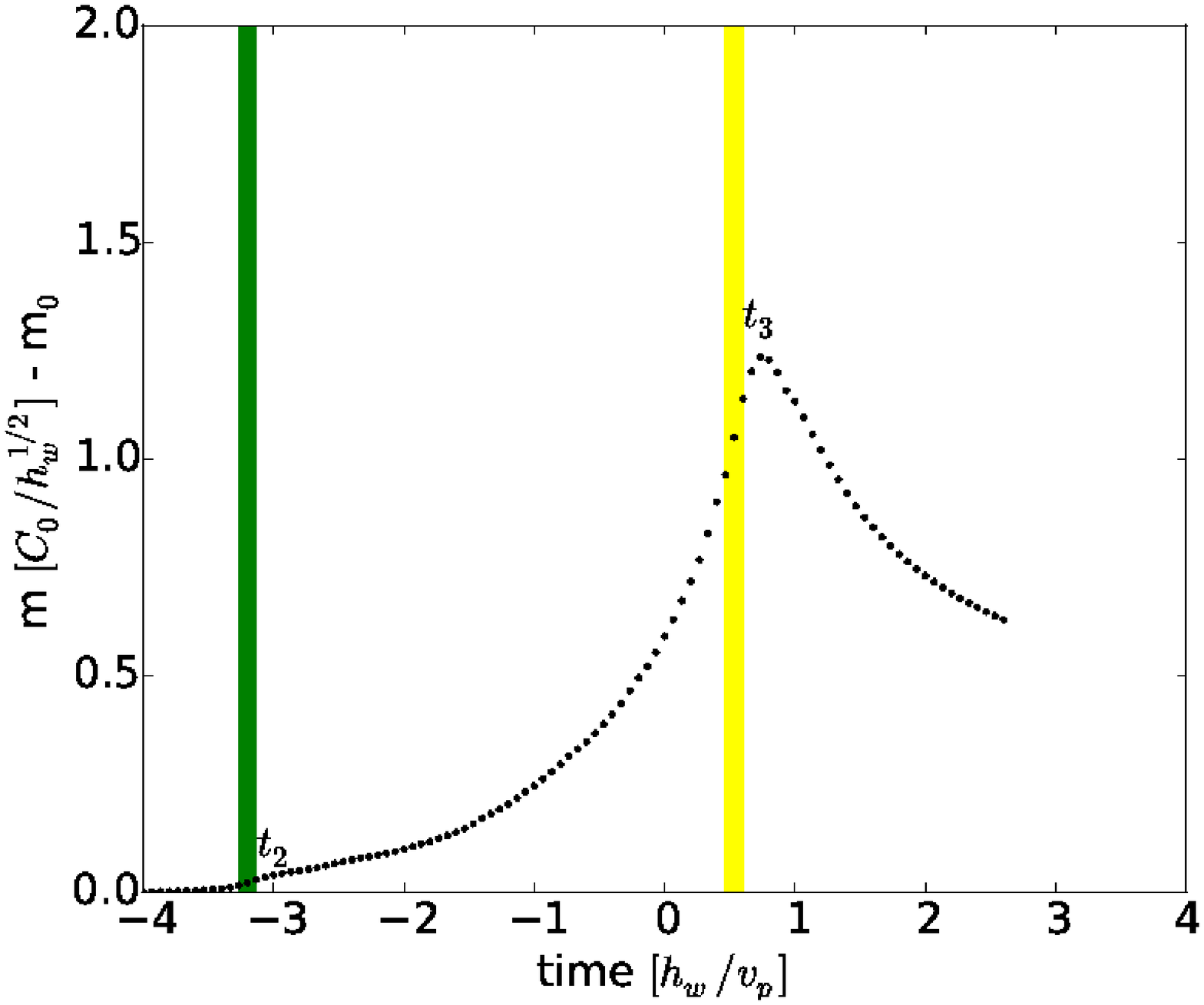}
\caption{\label{fig:M87_plots} The one dimensional profile obtained
  from the numerical integration along the ordonata axis of the source
  image is presented in the upper panel. The corresponding
  lightcurve associated to the 1D profile is presented in the lower
  panel. The points p2 and p3 mark the start and end of the equivalent
  dark disk and are associated with the moments of their overlap with
  the caustic: t2 and t3.}
\end{figure}

\section{Fitting mock data}\label{sec:numerics}

Having used the simple model of an ideal fold to gain insight, we now
move to the numerical computation of microlensing light curves with
realistic mass distributions, and then fit source models to
lightcurves in the presence of random and systematic errors.

First we generate microlensing magnification maps using the
microlensing code created by Joachim Wambsganss \citep{1999A&A...346L...5W, 
1990PhDT.......180W, 1999JCoAM.109..353W}. We then generate
mock light curves for three source models --- a crescent, a uniform
disc and a Gaussian disc --- and then fit each mock light curve to all
three source models. Parameter fitting and marginalising was done by
Markov-chain Monte Carlo (MCMC).

\subsection{Numerical microlensing magnification maps}

The code uses a ray-shooting technique to compute the gravitational
lensing effect of a mass distribution consisting of (a)~a smooth
component and (b)~a random distribution of point masses representing
stars.  The ray shooting maps a grid of $\vec\theta$ to $\vec\beta$
using the lens equation (\ref{eqn:lens}).  That is, the rays are shot
from the observer back to the source.

For the computation of the individual deflection angles
$\vec\alpha(\vec\theta)$ a hierarchical or tree method is used.  The
positions of all lensing masses are put into a grid of $\vec\theta$.
Each grid cell is subdivided into four smaller squares recursively
until every cell contains only one mass.  Nearby masses are added
individually while distance masses are clumped into larger grid cells
whose net contribution is approximated by its first few multipole
moments.  Scaled units are used, with the constant pre-factor in the
deflection angle (\ref{eqn:alpha}) separated out.

The result of ray shooting is a pixel map on the $\vec\beta$ plane of
the number of lightrays which arrive at the source plane from a
particular observer.  This intermediate result is effectively a
magnification map on the source plane.  Once the map is created, the
lightcurve can be obtained by specifying the transit path of the
source across the map.  At each point on this transit line, the code
computes equation (\ref{eqn:ft2d}) convolving the brightness
distribution of the source with the magnification pattern of the map.
In real life, not only are both lens and source moving but the lens
lens configuration, and with it the magnification pattern, is also
changing with time.  While the first subtlety is taken care of by a
coordinate transformation in this analysis, for the second one the
lens configuration is assumed to be constant in time.

When generating the magnification map depicted in
Figure~\ref{fig:magnification_map}, which is used in our analysis
below, only two point masses were included.  This was done in order to
have clean fold caustics.  For the computation of the actual
lightcurve, the code was modified to also allow for crescent shaped
images specified through the parameter set $R_p,R_n,a,b$. Here $R_p$
denotes the outer radius of the crescent and $R_n$ the inner one. The
orientation of the source image with respect to the magnification map
is specified by the parameters $a$ and $b$ as the shift of the center
of the inner disk from the center of the outer disk in $x$- and
$y$-direction respectively. In the original version of the code,
gaussian and disk-shaped images were already implemented. Those are
completely characterised by the single parameter $R_p$. The values of
the parameters are specified in pixel units corresponding to the
magnification map. Further, one needs to specify the start and end
point coordinates of the path, which the center of the source image
follows through the magnification map (see the depiction in figure
\ref{fig:magnification_map}). Hence, the points along this path are
specified through the number of time steps for which the computation of
the brightness is to be carried out. Those points correspond to the
actual measurement of the brightness of an object in the observational
case. For each timestep the two-dimensional convolution of equation
\ref{eqn:ft2d} is carried out numerically for the position of the
source on the magnification map.

For the purpose of this analysis, it was desirable to mimic the
analytical behaviour of a simple fold as much as possible, for
comparing the numerical result with the analytical one, therefore, the
path of the source was chosen, so that it intersects the border of the
caustic perpendicularly, and on a point where the border is a fold
caustic.

\begin{figure}
\centering
\includegraphics[width=0.9\hsize]{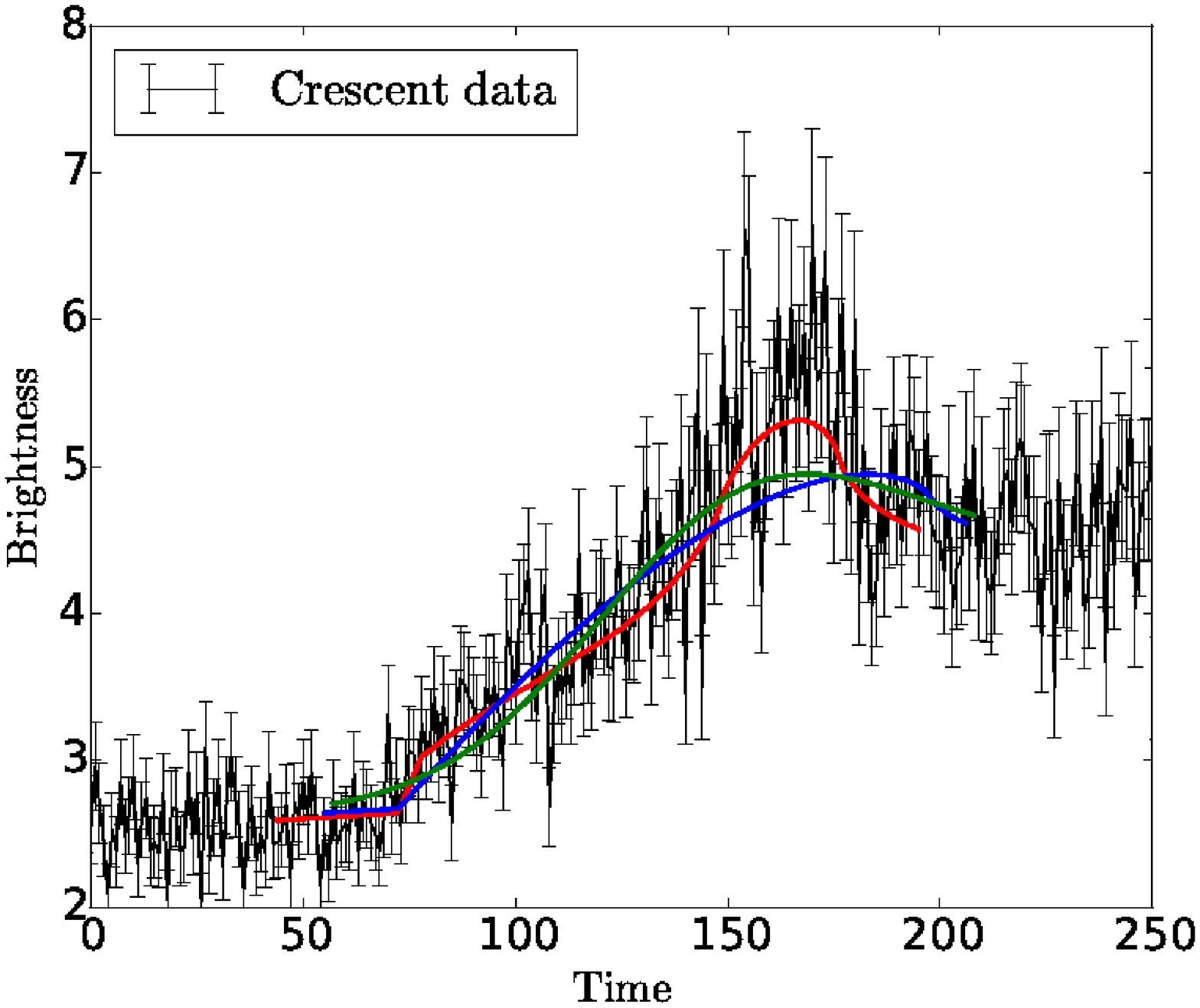}
\includegraphics[width=0.9\hsize]{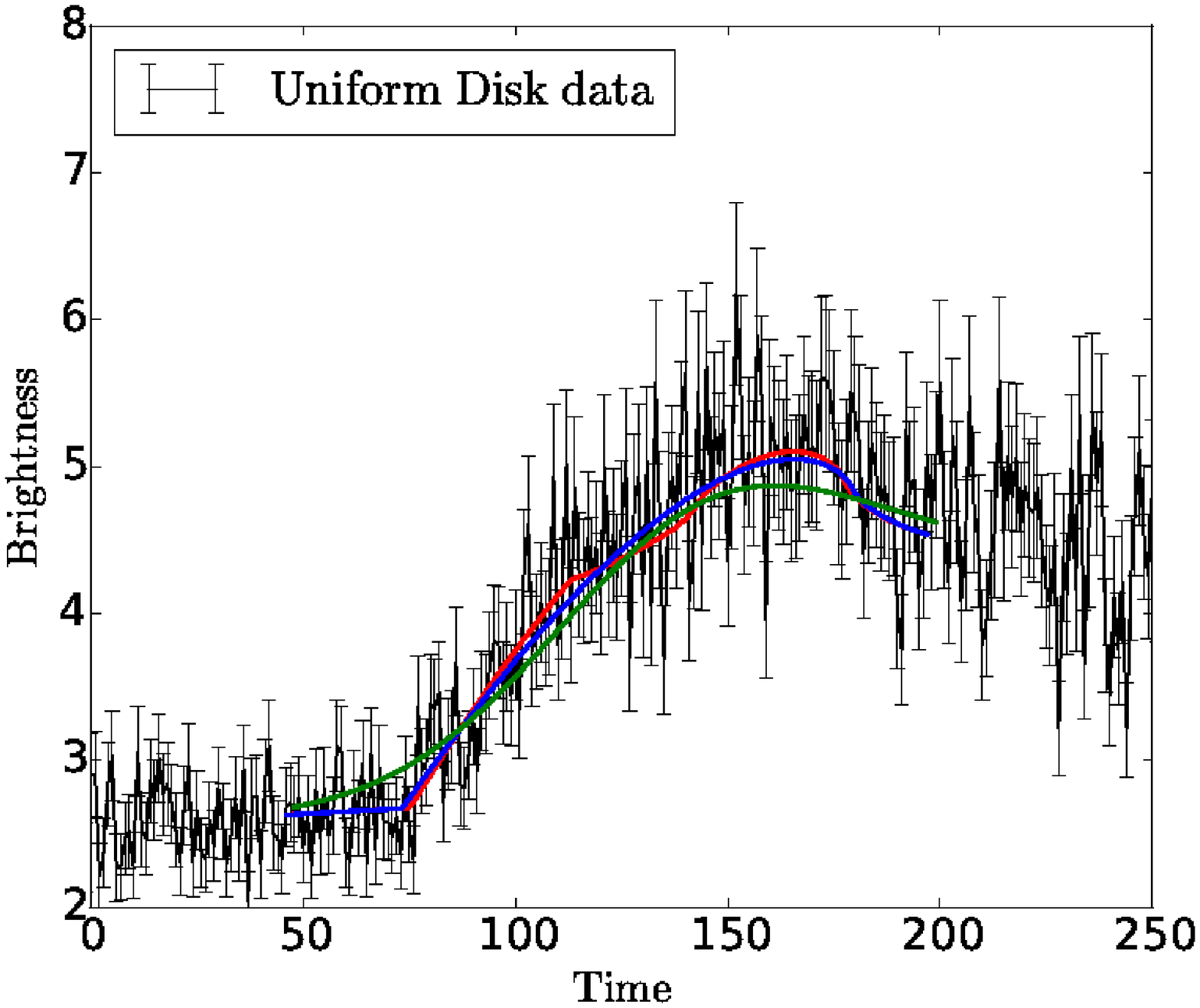}
\includegraphics[width=0.9\hsize]{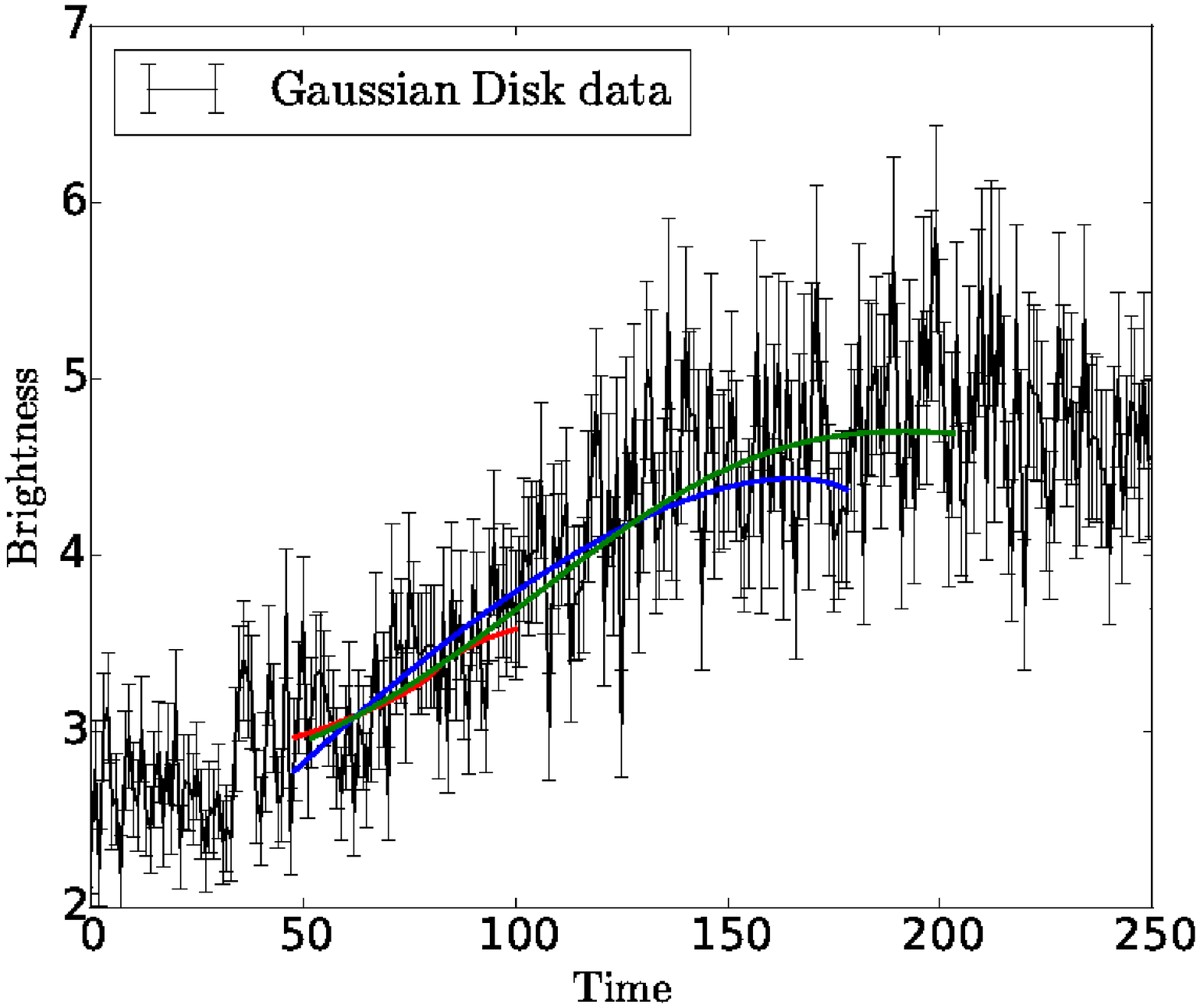}
\caption{\label{fig:mockdata} Three different mock datasets with
  three different parametrizations fitted to each --- crescent (red),
  uniform-disk (blue) and Gaussian disk (green).}
\end{figure}

\subsection{Mock light curves}

Mock light curves were generated by running a source across a
caustic, specifically along the line from point C to point B in
Figure~\ref{fig:magnification_map}.  The model light curves, to which
these are fitted, are made by running a source across a different
caustic, along the line from point A to point B in the same figure.
That is, the mock data are generated with a clean but not ideal fold
caustic and then fitted with another such caustic.  This mimics the
unavoidable systematic error of not knowing the caustic exactly.

The crescent source had parameters (as explained in
\S\ref{subsec:crescent}) $R_p$ and $R_n$ being the radii of the outer
and inner circles and $(\alpha,\beta)$ being the coordinates of the
inner circle with respect to the centre of the outer circle in the
coordinate system associated with the image in Figure 1 and not
associated with the caustic surface as defined in the previous
sections.  The parameter values were
\begin{equation}
   (R_p, R_n, \alpha, \beta) = (50.0, 30.0, 15.0, 10.0).
\label{eqn:cp}
\end{equation}
The uniform disc had the same (outer) radius as the crescent.  In the
Gaussian source, we set $3\sigma=50$.  For convenience, below we will
refer to $R_p$ of the Gaussian source, by which we actually mean
$3\sigma$.

Each mock light curve also has three nuisance parameters, namely the
beginning and end of the event and the brightness normalisation.
These are to be marginalised out by the MCMC.

Figure~\ref{fig:mockdata} shows the three light curves.  Each has 250
points regularly spaced in time, with Gaussian noise at the level of
10\% of the current brightness.  This amounts formally to a single
summary data point with a signal to noise of
$10\times\sqrt{250}\simeq160$.

\subsection{Model fitting and likelihood analysis}

Each of the three mock light curves was fitted to all three models.
We denote the nine possible cases with letter pairs, with the first
letter denoting the assumed model for the fitting procedure and 
the second letter representing
the source: thus CG means a {\em C\/}rescent model was fitted to mock
data from a {\em G\/}aussian source, DC stands for a uniform {\em
  D\/}isc model fitted to mock data from a {\em C\/}rescent source,
and so on.

\begin{figure}
\centering
  \includegraphics[width=0.9\hsize]{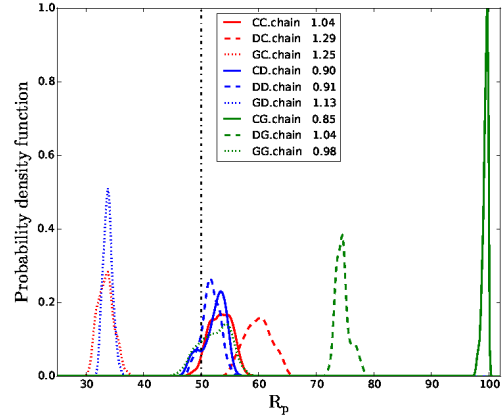}
\caption{\label{fig:mcmc} Posterior probability distribution for the
  source size $R_p$.  The vertical line is the correct value. Same
  color represents the same dataset whereas same line style
  corresponds to same model fit.  The legend gives the reduced
  $\chi^2$ of the best fit in each case.}
\end{figure}

Figure~\ref{fig:mcmc} shows the posterior probability distribution of
$R_p$ for all nine cases, along with the minimum reduced $\chi^2$ for
each case.  The area under each curve is unity. Note that the height
of the curves are not the likelihood, they are probability densities
in parameter space.

\begin{figure}
\centering
  \includegraphics[width=0.9\hsize]{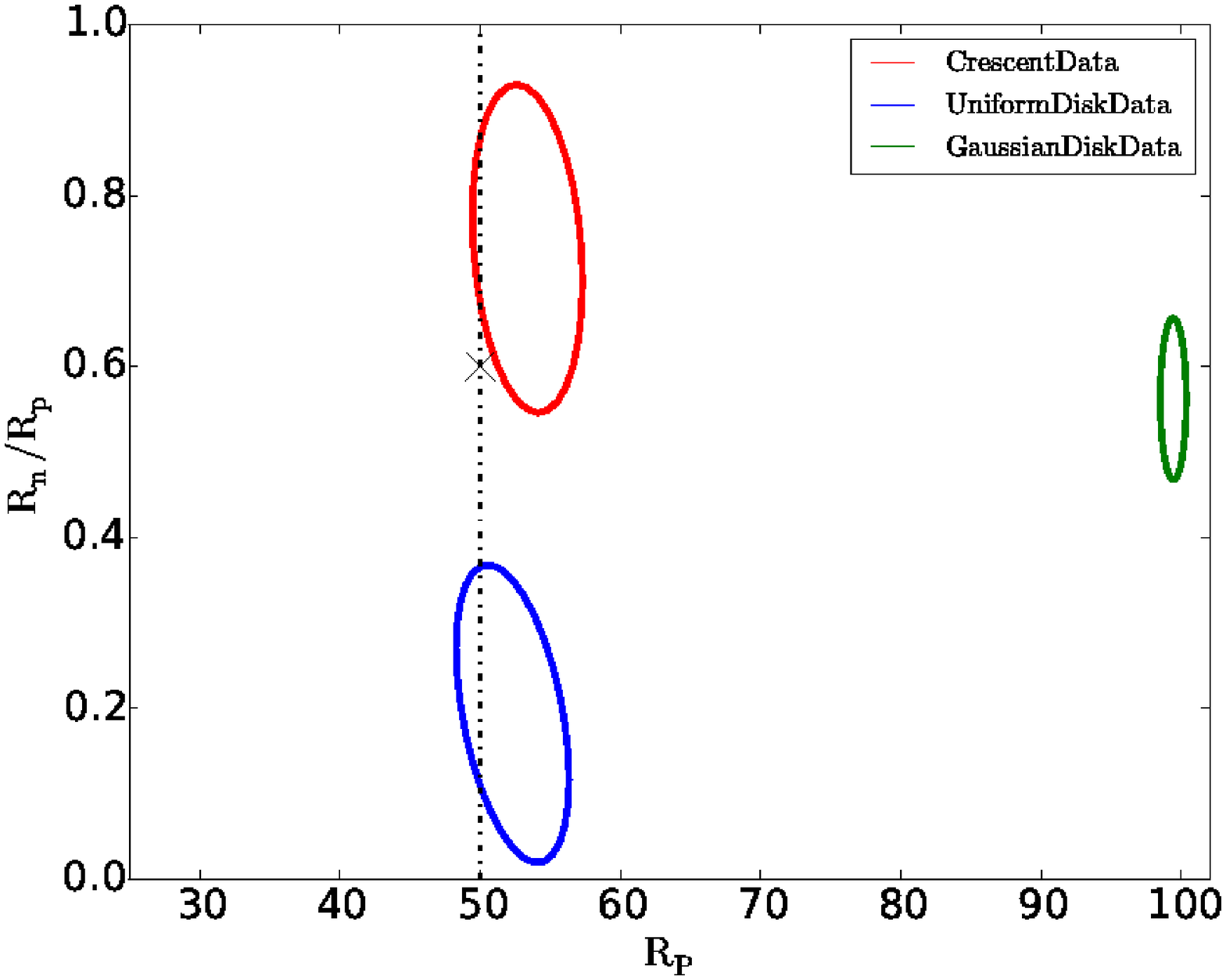}
  \includegraphics[width=0.9\hsize]{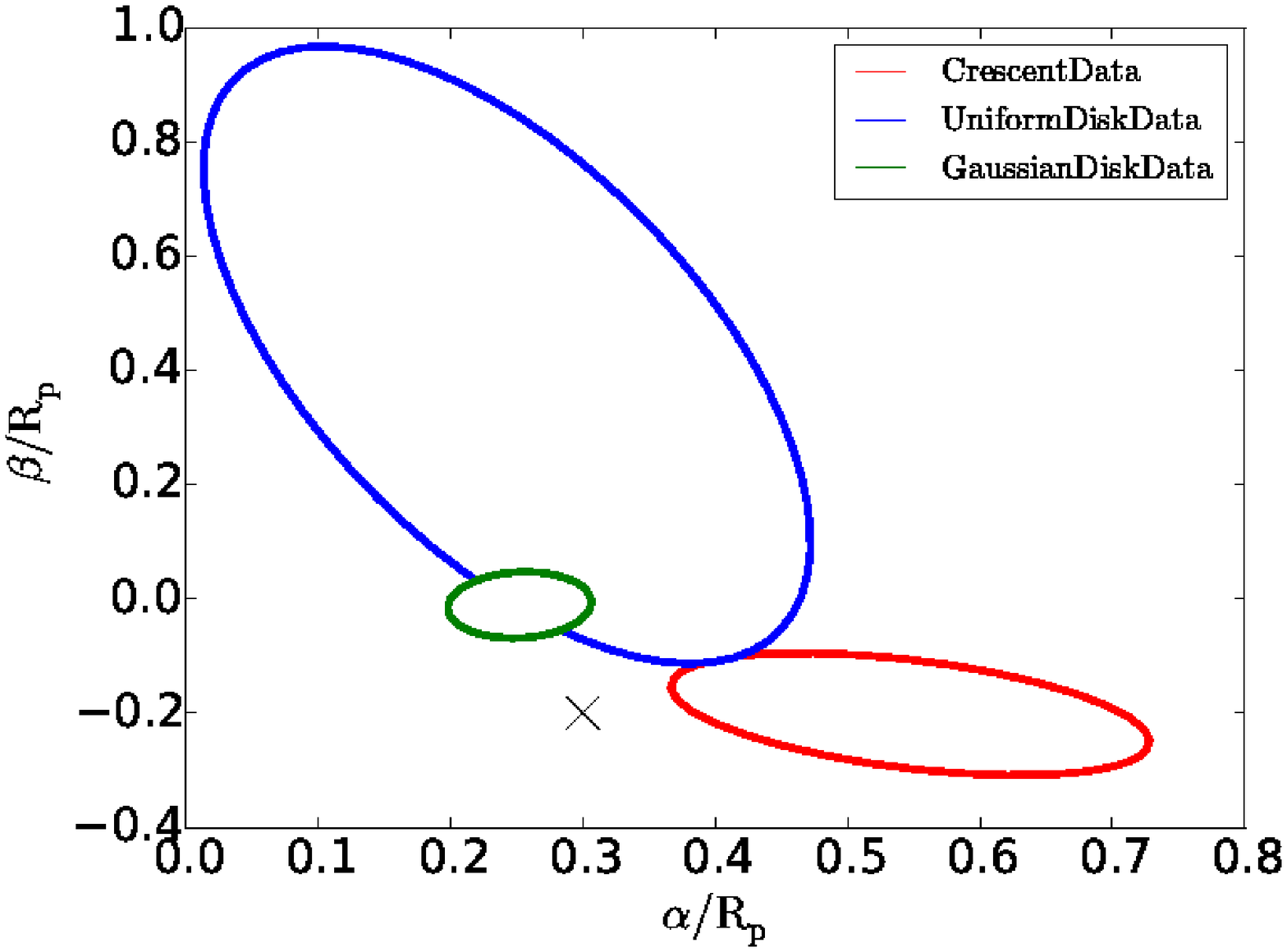}
\caption{\label{fig:crescentfit} The 2$\sigma$ contours (or error ellipses)
  of the crescent model when fitting with the three different
  datasets.}
\end{figure}

Let us first consider the three cases where a crescent model was
fitted.  These are the solid curves in Figure~\ref{fig:mcmc}, with the
colours of the curves indicating the source.  Meanwhile,
Figure~\ref{fig:crescentfit} shows $2\sigma$ of the inferred parameter
values.

\begin{enumerate}

\item[1) {\bf CC}:] solid red curve in Figure~\ref{fig:mcmc} and red
  ellipses in Figure~\ref{fig:crescentfit}.  In this case, a light
  curve from a crescent source was being fitted to a crescent model.
  The fit gives reduced $\chi^2$ close to unity, as expected.  The
  recovered parameter values are near or slightly outside the
  $2\sigma$ ellipses.  This is expected since we added a small
  systematic error, by using different caustics (though both clean
  folds) for the mock data and the model fit.  The apparent degeneracy
  between $a$ and $b$, seen in Figure~\ref{fig:crescentfit}, is also
  expected, since only the distance of the crescent's small circle
  from the caustic influences the magnification.

\item[2) {\bf CD}:] solid blue curve and blue ellipses.  Here a
  crescent model was fitted to a light curve from a uniform disc.  The
  situation is formally a crescent with $R_n=0$ and $a,b$ arbitrary,
  and this shows in the blue ellipses in the recovered parameter
  values.  The redundant parameters $\alpha$ and $\beta$, in effect, allow the
  model to partly fit the noise, and hence, the $\chi^2$ is somewhat
  lower than in the previous case.

\begin{figure}
\centering
\includegraphics[width=0.9\hsize]{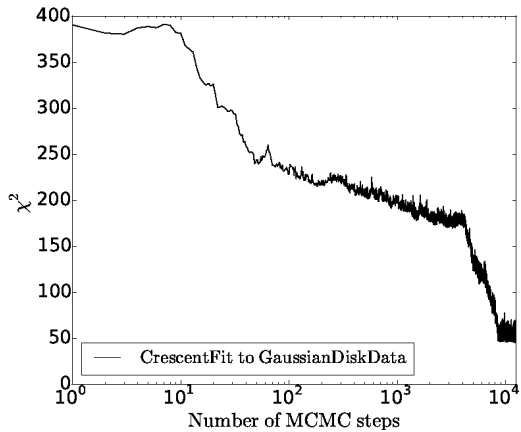}
\caption{\label{fig:burnin} The $\chi^2$ of each step of the full MCMC
  for one case.}
\end{figure}

\item[3) {\bf CG}:] Solid green curve and green ellipses.
  Additionally, Figure~\ref{fig:burnin} shows the progress of
  reduction of $\chi^2$ in this case, where a crescent-source model is
  used to fit data from a Gaussian source.  The parameter values are
  seemingly tightly constrained, but the recovered $R_p$ is completely
  wrong since the correct value is far from the green curve and green
  ellipses.  The reduced $\chi^2$ is only 0.85, indicative of
  over-fitting.  We can see what has happened from the red curve in
  the bottom panel of Figure~\ref{fig:mockdata}.  The best-fit model
  covers only a small part of the light curve.  That is, the fitting
  procedure has exploited the nuisance parameters to find a good fit
  to mainly noise.

\end{enumerate}

Next we consider the three cases where a disc model is fitted.  These
correspond to the dashed curves in Figure~\ref{fig:mcmc}.  There is
only one interesting parameter to fit, the disc radius $R_p$, and no
equivalent of Figure~\ref{fig:crescentfit} is needed. 

\begin{enumerate}

\item[4) {\bf DC}:] red dashed curve.  On fitting a disc model to a
  crescent-source source, the recovered $R_p$ is incorrect, but the
  reduced $\chi^2$ of 1.29 signals that the data reject the model.

\item[5) {\bf DD}:] blue dashed curve.  When the correct model is
  fitted to a disc-source, the best fit $\chi^2=0.91$ is good and
  $R_p$ is recovered within the uncertainty estimate.

\item[6) {\bf DG}:] green dashed curve.  On fitting a disc model to a
  Gaussian-source model, the recovered $R_p$ is incorrect, but the
  reduced $\chi^2$ gives no signal that something is amiss.  It
  appears that a Gaussian source could mimic a uniform disc.

\end{enumerate}

Finally, we consider the three cases where a Gaussian model is fitted.
These correspond to the dotted curves in Figure~\ref{fig:mcmc}.  Again,
there is only one interesting parameter to fit, the Gaussian
$3\sigma$-radius, which we have called $R_p$.

\begin{enumerate}

\item[7) {\bf GC}:] red dotted curve.  When a Gaussian-source model is
  fitted to data from a crescent source, the recovered $R_p$ is wrong
  but the reduced $\chi^2=1.29$ shows the data rejecting the model.

\item[8) {\bf GD}:] blue dotted curve.  When a Gaussian-source model
  is fitted to data from a uniform disc, the recovered $R_p$ is wrong
  but the model is rejected anyway.

\item[9) {\bf GG}:] green dotted curve.  When data from a Gaussian is
  fitted to the correct model, $R_p$ is recovered within it's estimated
  uncertainty and the best fit reduced $\chi^2$ is close to unity.

\end{enumerate}

The above results suggest the following strategy for fitting a
lightcurve from an unknown source profile: first try a Gaussian-source
model; if the data reject that model, try a uniform disc; if the
uniform disc is also rejected by the data, try a crescent model.  Our
numerical experiments indicate --- assuming one of the three sources
models are correct --- that the reduced $\chi^2$ would unmask the
correct model, and its parameters would be correctly recovered.

\section{Discussion}\label{sec:discussion}

In the current paper, we simulate and study the resulting microlensing lightcurves of geometric crescent-shaped sources 
and compare them with the microlensing lightcurves of other simple mathematically describable source profiles. 
In order to mimic the behaviour of the flux of light from the source in the proximity of a fold caustic, we make use of the simple approximation described in equation (5). 
The equation would exhaustively describe the magnification map and offer a good universal approximation for the 
particular microlensing regime that we consider. 
Namely, the shape of the caustic boundary in the proximity of the source in the respective plane can be approximated by
 a line due to reason that the local radius of curvature of caustic is orders of magnitude greater than the half-light
 radius of the studied source. 
In particular cases in which the previously mentioned approximation loses its validity, the impact on the quality of the 
lightcurves is not evenly distributed. The shape of the lightcurve will be maintained. 
The data points corresponding to the source position before and during the overlapping of the caustic will be affected by
 smaller errors than the data points corresponding to later times. \\

The first two source profiles that we consider are the uniform disk and symmetric Gaussian source. 
Both of them can be described by a half-light radius $r_{1/2}$ and a total unlensed light flux $S_0$. 
With the two parameters constrained the one-dimensional profiles, as well as the lightcurves of the two source, are completely determined, since no free parameter remains. 

The previous statement does not hold for a crescent source.
In the case of the crescent source there are in total five parameters: the integrated flux of the source $S_0$,
 the radii of the bright/dark disk $R_p$/$R_n$ and the displacement of the centers of the two disks on the axes perpendicular and parallel to the caustic $a$ and $b$. 

Two of the parameters can be reduced by expressing the results in terms of $S_0$ and $r_{1/2}$. 
The later being determinable for any set of parameters $R_p, R_n$ and $a^2+b^2$. Moreover, one of the displacement 
parameters $b$ has no impact on the one-dimensional profile of the source that results from the projection of the source image on an axis perpendicular to the caustic. 

Since the one-dimensional source profile that corresponds to an axis perpendicular to the caustic contains exhaustively all the information regarding the source that can be revealed by the lightcurve, 
the value of the parameter $b$ does not have an effect on the shape of the lightcurve. 

Nevertheless, the $b$ parameter is relevant for the calculation of $r_{1/2}$. 
It's qualitative effect is to decrease the value of the half-light radius when the absolute value of the parameter is increased.  
With two parameters constrained and another irrelevant to the shape of the lightcurve, two free parameters remain $R_n$ and $a$. 
Figure 4 reveals that the lightcurve of a crescent source has more visible features than the other two light-curves 
corresponding to the disc and Gaussian shape. The parameters $R_n$ and $a$ have strong influences on the shape of 
the microlensed lightcurve as can be seen in figures 5, 6 and 7. Moreover, the one-dimensional source profile corresponding to the direction perpendicular to the caustic reveal four characteristic points. The overlap of each of these points with the caustic leaves visible features on the lightcurve at the corresponding instances of time. In timely order, the instances correspond to the start of the overlap between the caustic and the bright disk, the start of the overlap between the caustic and the dark disk, the end of the overlap between the caustic and dark disk and finally the end of the overlap between the caustic and the bright disk.

With the different source profiles and their corresponding lightcurves studied we can change our point of view of the system to that of an observer. 
The observer would basically detect only the lightcurve of such a source. As described in section 4.3 the timing of the onset and offset of the previously 
described periods can be used to estimate the values of the radii and one of the displacement parameters when assuming a geometric crescent. 
All quantities can be estimated in terms of the relative velocity of the source in a direction perpendicular to the caustic.

Furthermore, a simulated image of M87 presented in \citep{2012MNRAS.421.1517D} has been microlensed (figure 9). On the resulting lightcurve the instances 
corresponding to the start and end of the black hole shadow and caustic overlap were distinguishable.

In the case of a high-quality lightcurve with insignificant noise and measurement errors, the parameters can be obtained 
by simply identifying the characteristic instances of time without making use of the actual values of the magnification
 map. If the effect of the errors and the noise distorts the magnification time function enough so that the 
characteristic epochs are not identifiable with the characteristic periods still visible, the boundaries of the periods 
can be roughly estimated. Furthermore, if direct estimates of the parameters cannot be obtained we propose the use of a strong
statistical tool such as Markov-Chain Monte Carlo. In Section 6  we have studied the possibility of identifying a crescent 
source and  the possibility of recovering the respective parameters. As magnification map, we have used a complex numerical 
one generated with the microlensing code by \cite{1999A&A...346L...5W}. In addition, we have added to the signal a Gaussian 
noise with an SNR of 1.6.  In the experiment, we have considered all nine combinations of original source profiles and assumed fitting source models.
Effectively we have fitted using MCMC all three sources with all three assumed fitting models. The results of the experiment allowed us to build a 
procedure for distinguishing the shape of the source assuming that one of the three models we have considered is a good approximation.  
The procedure would be useful for observers that endeavour to gain more information about an unresolved source for which they can study 
the microlensing lightcurve.  As a first step in the procedure, one should first attempt to use MCMC with a Gaussian model assumption. 
If either the value of the $\chi^2$ or the number of rejected datapoints is large then the next step is to change the assumed source model to 
a uniform disc and redo the fitting. Finally, if the uniform disc assumption is rejected as well then the fitting should be done with a 
crescent source model assumption. If at each of the three steps the data rejects the model then the source cannot be approximated by any of the three models. 
Otherwise, if at one of the steps the data does not reject the model then the respective model is a good approximation.   

The previously mentioned abstract parameters can be related to
physical quantities specific to the central region of a quasar.  As
such the luminous region would correspond to the bright accretion disc
that surrounds the black hole. The later's gravity would cause a
shadow in the bright region limited by the extent of the event horizon
of the black hole. Therefore, the radius of the bright disk would
provide an estimate of the size of the accretion disk and the radius
of the dark disc would provide and estimate of the gravitationally
magnified Schwarzschild radius of the black hole $R_{S}^{magnified} =
\Delta t_{dark} \cdot v_p$.  By $R_{S}^{magnified}$ we refer to the
apparent Schwarzschild radius which is larger than the real value at
large distances due to the black hole's own gravity.  Moreover, the
gravitationally magnified value of the Schwarzschild radius is a
monotonic function of the black hole's mass. Therefore, it can be used
to estimate the mass of the black hole if it was not rotating. In the
previous expression, the $\Delta t_{dark}$ denotes the period of
overlap between the black hole shadow and fold caustic.  $v_p$ denotes
the component of the relative velocity of the source and fold which is
perpendicular to the caustic.  The respective velocity is an unknown,
though it can be constrained on a case by case basis to an order of
magnitude or even better. This would require the study of the dynamics
of the stellar structure which contains the gravitational lens.  A
better estimate of the relative velocity would facilitate a better
estimate of the effective non-rotating black hole mass associated with
the black hole shadow.

The parameters whose values cannot be determined due to the loss of information from the directions parallel to the 
caustic could be obtained in the eventuality in which the same source crosses multiple caustics that are not parallel. 
Multiple crossing of caustics can reveal details of the one-dimensional flux profile corresponding to multiple distinct 
directions which would allow the reconstruction of the two-dimensional profile analogous to the process through which an image of a CT scan is obtained.  

\textit{Author contributions}\\
Prasenjit Saha provided the original idea and plan for the research project as well as multiple contributions to the analysis 
and manuscript preparation. Mihai Tomozeiu simulated and studied the ideal behaviour of the microlensing lightcurves for the 
different source profiles discussed and prepared the manuscript.
Joachim Wambsganss contributed to the research planning and provided the numerical code used by Manuel Rabold to create 
the magnification map and corresponding lightcurves used in the MCMC analysis performed by Irshad Mohammed in the last 
part of the presented work. Both Manuel Rabold and Irshad Mohammed had large contributions in writing the "Fitting Mock Data" section.

\section{Acknowledgement}
IM is supported by Fermi Research Alliance, LLC under Contract No. De-AC02-07CH11359 with the United States Department of Energy.
J.W. would like to acknowledge and thank the Pauli Center for Theoretical Studies of ETH Zurich and University of Zurich for 
generous support during the Schrödinger visiting professorship in 2013.

\bibliographystyle{mn2e}

\def\apj{ApJ}
\def\apjl{ApJL}
\def\aj{AJ}
\def\mnras{MNRAS}
\def\aap{A\&A}
\def\nat{nature}
\def\araa{ARAA}
\def\pasa{PASA}
\bibliography{heap}

\end{document}